\documentclass[aoas,MSNbibl,nameyear,dvips]{arximspdf}
\usepackage{algorithm}
\usepackage{algorithmicx}
\usepackage{algpseudocode}
\usepackage{graphicx}
\usepackage{url,breakurl}

\doi{10.1214/14-AOAS761} 
\volume{8}
\issue{3}
\pubyear{2014}
\firstpage{1892}
\lastpage{1919}
\docsubty{FLA}

\makeatletter
\def\aaa{\mbox{\fontsize{8.36}{10.36}\selectfont{-}}}
\newcommand{\rrVert}{\Vert}
\newcommand{\llVert}{\Vert}

\newproclaim{assumption}{Assumption}
\makeatother

\begin{document}
\begin{frontmatter}

\title{Joint estimation of multiple related biological~networks\thanksref{T1}}
\runtitle{Joint network inference}

\begin{aug}
\author[A]{\fnms{Chris J.}~\snm{Oates}\corref{}\ead[label=e1]{c.oates@warwick.ac.uk}\thanksref{m1}},
\author[B]{\fnms{Jim}~\snm{Korkola}\ead[label=e2]{korkola@ohsu.edu}\thanksref{m2}},
\author[B]{\fnms{Joe~W.}~\snm{Gray}\ead[label=e3]{grayjo@ohsu.edu}\thanksref{m2}}
\and
\author[C]{\fnms{Sach}~\snm{Mukherjee}\ead[label=e4]{sach@mrc-bsu.cam.ac.uk}\thanksref{m3,m4,T2}}
\runauthor{Oates, Korkola, Gray and  Mukherjee}
\affiliation{University of Warwick\thanksmark{m1}, OHSU Knight Cancer
Institute\thanksmark{m2}, MRC Biostatistics Unit\thanksmark{m3} and University of Cambridge\thanksmark{m4}}
\address[A]{C. J. Oates\\
Department of Statistics\\
University of Warwick\\
Coventry, CV4 7AL\\
United Kingdom\\
\printead{e1}} 
\address[B]{J. Korkola\\
J. W. Gray\\
Knight Cancer Institute\\
Oregon Health and Science University\\
Portland, Oregon 97239\\
USA\\
\printead{e2}\\
\phantom{\textsc{E-mail:}\ }\printead*{e3}}
\address[C]{S. Mukherjee\\
MRC Biostatistics Unit and\\
School of Clinical Medicine\\
University of Cambridge\\
Cambridge, CB2 0SR\\
United Kingdom\\
\printead{e4}}
\end{aug}

\thankstext{T1}{Supported in part by NCI U54 CA112970, UK EPSRC
EP/E501311/1 and EP/D002060/1, and the Cancer Systems Biology Center
grant from the Netherlands Organisation for Scientific Research.}
\thankstext{T2}{A recipient of a Royal Society Wolfson Research Merit Award.}

\received{\smonth{2} \syear{2013}}
\revised{\smonth{4} \syear{2014}}

\begin{abstract}
Graphical models are widely used to make inferences concerning
interplay in multivariate systems. In many applications, data are
collected from multiple related but nonidentical units whose underlying
networks may differ but are likely to share features. Here we present a
hierarchical Bayesian formulation for joint estimation of multiple
networks in this nonidentically distributed setting. The approach is
general: given a suitable class of graphical models,
it uses an exchangeability assumption on networks to provide a
corresponding joint formulation. Motivated by emerging experimental
designs in molecular biology, we focus on time-course data with
interventions, using dynamic Bayesian networks as the graphical models.
We introduce a computationally efficient, deterministic algorithm for
exact joint inference in this setting.
We provide an upper bound on the gains that joint estimation offers
relative to separate estimation for each network and empirical results
that support and extend the theory, including an extensive simulation
study and an application to proteomic data from human cancer cell
lines. Finally, we describe approximations that are still more
computationally efficient than the exact algorithm and that also
demonstrate good empirical performance.
\end{abstract}

\end{frontmatter}

\section{Introduction}

Graphical models are widely used to represent multivariate systems.
Vertices in a graph (or network; we use both terms interchangeably)
$G$~are identified with random variables and edges between the vertices
describe conditional independence statements or, with suitable modeling
and semantic extensions, causal influences between the variables.
In many applications a key statistical challenge is to construct a
network estimator $\hat{G}(\mathbf{y})$, based on data $\mathbf{y}$, that
\mbox{performs} well in a sense appropriate to the application.
Such ``network inference'' is increasingly a mainstream approach in many
disciplines, including neuroscience, sociology and computational biology.

Network inference methods usually assume that the data are identically
distributed (specifically, that data sets satisfy an exchangeability
assumption).
However, in many applications, data are not identically distributed,
but are instead obtained from multiple related but nonidentical units
(or ``individuals''; we use both terms interchangeably). This paper
concerns network inference in this nonidentically distributed setting.

Our work is motivated by biological networks in cancer.
Multiple studies have demonstrated the remarkable genomic heterogeneity
of cancer [\citet{Thousand,TCGA}].
At the same time, the question of how such heterogeneity is manifested
at the level of biological networks has remained poorly understood.
We focus in particular on protein signaling networks in human cancer
cell lines. Signaling networks describe biochemical interplay
between
proteins and are central to cancer biology.
However, sequence and transcript data alone are inadequate for the
study of signaling and, indeed, these data types can be discordant with
the abundance of signaling proteins and post-transitional modifications
(including phosphorylation) that are key to the process [\citet{Akbani}].
Recent developments in proteomics, including reverse-phase protein
arrays [or RPPA, see \citet{Hennessy};
this technology provides the data we analyze below],
have improved the ability to interrogate signaling heterogeneity.

To fix ideas, we begin by describing the specific application that
motivates this work.
We consider time-course phosphoprotein measurements obtained using RPPA
technology (details appear below) for 6 cell lines.
The goal of the study is to infer cell line-specific protein signaling
networks $G^j$, $j = 1,\ldots,6$, and additionally to highlight
experimentally testable differences between them.
Prior network information is available from the literature, but it is
believed that cell line-specific genetic alterations may induce
differences with respect to the ``literature network'' (and between cell
lines). At the same time, the amount of data per cell line is limited
(6 time points in each of 4 conditions, making a total of 24 data
points per cell line $j$, constituting data $\mathbf{y}^j$). Since the
cell lines $j$ are closely related, yet potentially different with
respect to underlying networks, a key inferential question is how to
``borrow strength'' between the network estimation problems.\vspace{1pt} That is, we
seek a joint estimator of the cell line-specific networks $\{ G^1
\cdots G^6 \}$ based on the entire (nonidentically distributed) data
set $\{\mathbf{y}^1 \cdots\mathbf{y}^6 \}$ that shares information
between the estimation problems while preserving the ability to
identify cell line-specific network structure.

This application is an example of a more general class of biological
applications, where individuals $j$ could correspond to, for example,
different patients or cell lines (or groups thereof; e.g., disease
subtypes) and the networks themselves to gene regulatory or protein
signaling networks that could depend on the genetic and epigenetic
state of the individuals.
Indeed, continuing reduction in the unit cost of biochemical assays has
led to an increase in experimental designs that include panels of
potentially heterogeneous individuals [\citet{Barretina,Cao,Maher,TCGA}]. As in the signaling example above, in such
settings, given individual-specific data $\mathbf{y}^j$, there is
scientific interest in individual-specific networks $G^j$ and their
similarities and differences.

Following \citet{Werhli,Penfold} and others, we focus on the case of
directed networks $G^j$ that are exchangeable in the sense that
inference is invariant to permutation of individuals $j \in\mathcal{J}
= \{1,\ldots,J\}$.
We model data on all individuals $\{\mathbf{y}^j \dvtx j \in\mathcal{J}\}$
within a joint Bayesian framework.
Regularization of individual networks is achieved by introducing a
latent network $G$ to couple inference across all individuals. We
report posterior marginal inclusion probabilities for every possible
edge in each individual network $G^j$ as well as the latent network $G$.
The high-level formulation we propose is general and, in principle,
essentially any graphical model of interest could be embedded within
the framework described to enable joint estimation.

\begin{figure}[b]

\includegraphics{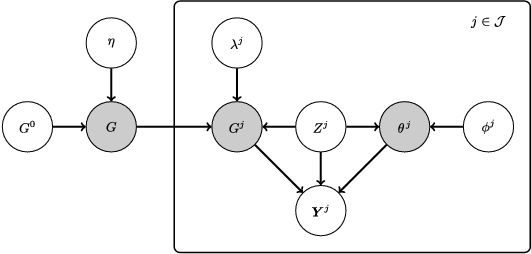}

\caption{Joint network inference (JNI). A hierarchical model for
analysis of multivariate data from multiple, nonidentical units or
individuals, indexed by $j$. (Shaded nodes are unobserved. $G^0={}$prior
network, $G={}$latent network, $G^j={}$network\vspace{1pt} specific to individual
$j$, $\bolds{\theta}^j={}$parameters for individual $j$, $\mathbf{Y}^j={}$observables for individual $j$, $Z^j={}$ancillary\vspace{1pt} information available
on individual $j$, $\eta,\lambda^j={}$inverse temperature
hyperparameters, $\phi^j={}$hyperparameters defining a prior on $\bolds{\theta}^j$. Panel notation is used to indicate the presence of
multiple individuals $j \in\mathcal{J}$.
Note that in practice we take $\lambda_j \equiv\lambda$ for all $j \in
\mathcal{J}$.)}
\label{model}
\end{figure}

In general, the individual $j$'s could have complex, hierarchical
relationships, for example, with $j$'s belonging to groups and
subgroups [e.g., corresponding to cancer types and subtypes; see \citet{Curtis}].
The exchangeable case we consider corresponds in a sense to the
simplest possible hierarchy in which each individual is dependent on a
single latent graph (see Figure~\ref{model}).
In settings where groups can be treated as approximately homogeneous,
the approach presented in this paper can be trivially used to give
group-level estimates, by using
$j$ to index groups rather than individuals, with all data for group
$j$ modeled as dependent on graph $G^j$. This corresponds to an
assumption of identically distributed data within (but not between) groups.
In the empirical study presented below we consider also robustness of
our approach under violation of the exchangeability assumption.

For the application to time-course data from protein signaling that we
focus on, we present a detailed development using directed graphical
models called dynamic Bayesian networks (DBNs).
These are directed acyclic graphs (DAGs) with explicit time indices
[\citet{Murphy}].
The main contributions of this paper are as follows:

\begin{itemize}
\item \textit{Bayesian computation}. For the time-course setting, we put
forward an efficient and exact algorithm. This is done by exploiting
factorization properties of the DBN likelihood, analytic
marginalization over continuous parameters and belief propagation. In
moderate dimensional settings this allows exact joint estimation to be
carried out in seconds to minutes (we discuss computational complexity below).

\item \textit{Theory}. We provide a result that quantifies the statistical
efficiency of joint relative to separate estimation and that gives a
sufficient condition for improved performance.

\item \textit{Empirical investigation}. The availability of an efficient
Bayesian algorithm enables, for the first time, a comprehensive
empirical study of joint estimation, including a wide range of
simulation regimes and an application to experimental data from a panel
of human cancer cell lines.
For several
classical (nonjoint) DBNs, including a recent causal variant suitable
for interventional data [\citet{Spencer}], we formulate corresponding
joint estimators. This allows us to investigate the effect of joint
estimation itself;
we find that it often provides gains relative to the corresponding
individual-level estimators.
Some computationally favorable approximations to joint inference are
described that we find perform well under a range of conditions.
\end{itemize}

Joint estimation has previously been discussed in the Gaussian
graphical model (GGM) literature [\citet{Danaher}].
In contrast to GGMs, motivated by biological applications, we focus on
DAG models with a causal interpretation.
Approaches to context-specific DAG structure based on the embellishment
of Bayesian networks include \citet{Boutilier,Geiger}.
Our approach differs by regularizing based on network structure alone;
we do not place exchangeability assumptions on the data-generating parameters.
Related work that is based on DAGs includes \citet{Niculescu,Werhli,Dondelinger12}.
In a sequel to the present work, \citet{Oates6} provide an exact
algorithm for joint \textit{maximum a posteriori} (MAP) estimate of
multiple (static) DAGs. In contrast, here we focus on Bayesian
model-averaging (as opposed to MAP estimation) and on time-course data
(or, more generally, Bayesian networks with a fixed ordering of the variables).

In a similar vein to the present paper, \citet{Oyen} estimated multiple
DAGs sharing a common ordering of the vertices, but they considered
only applications involving $J=2$ individuals.
Our work is closely related to \citet{Penfold}, who also considered
Bayesian joint estimation of directed graphs from time-course data.
However, as we discuss in detail below, the methodology they propose is
prohibitively computationally expensive for the applications we
consider here.
In comparison, the exact algorithm we propose offers massive
computational gains that in turn allow us to present a much more
extensive study of joint estimation than has hitherto been possible.
Furthermore, we allow for prior information regarding the network
structure (including individual-specific characteristics) and present
theoretical results concerning the statistical efficiency of joint
network estimation.

The remainder of the paper is organized as follows. In Section~\ref{secmethods} we describe a hierarchical Bayesian formulation and in
Section~\ref{seccomputers} we discuss computationally efficient joint
inference in the case of DBNs.
Empirical results are presented in Section~\ref{secresults},
including an application to protein signaling in cancer. Finally, we
close with a discussion of our findings in Section~\ref{secdiscussion}.

\section{Joint network inference: The general case} \label{secmethods}

We describe a general statistical formulation for joint network
inference that can be coupled to essentially any class of graphical models.
For computational tractability it may be necessary to place
restrictions on the class of graphical models; in Section~\ref{seccomputers} we present a~detailed development for DBNs that are
well-suited to our motivating application in cancer.

\subsection{Hierarchical model} \label{secJNI}
Consider a space $\mathcal{G}$ of graphs on the vertex set $\mathcal{P}
= \{1,\ldots,P\}$.
To keep the presentation general, we do not specify the type of graph
or restrictions on $\mathcal{G}$ at this stage (the special case of
DBNs for time-course data is described below).
As shown in Figure~\ref{model}, each individual network $G^j \in
\mathcal{G}$ is modeled with dependence on a latent network $G \in
\mathcal{G}$ that in turn depends on a prior network $G^0 \in\mathcal
{G}$ (Section~\ref{priors}).
In this way, estimates of the individual networks $G^j$ are regularized
by shrinkage toward the common latent network $G$ that, in turn, may be
constrained by an informative network prior.
As in any graphical model, observations $\mathbf{Y}^j$ on individual $j$
are dependent upon a graph $G^j$ and parameters $\bolds{\theta}^j$.
Here $Z^j$ denotes any ancillary information available on individual $j$.
The model is specified by
%
\begin{eqnarray}
\label{Gibbs2}
p\bigl(G|G^0,\eta\bigr) & \propto& \exp \bigl(-\eta d
\bigl(G,G^0\bigr) \bigr),
\\
\label{Gibbs1}
p\bigl(G^j|G,\bolds{\lambda},Z^j\bigr) & \propto& \exp
\bigl(-\lambda^j d^j\bigl(G^j,G;Z^j
\bigr)\bigr)
\end{eqnarray}
and a suitably chosen graphical model likelihood $p(\mathbf{Y}^j|G^j,\bolds
{\theta}^j,Z^j)$.
Equation (\ref{Gibbs2}) follows the ``network prior'' approach of \citet{Mukherjee2} that was proposed for biological applications where
subjective prior structural information is available.
The functionals $d^j,d\dvtx\mathcal{G} \times\mathcal{G} \rightarrow\mathbb{R}$ and hyperparameters $\eta,\lambda^j$ must be specified (Section~\ref{priors}).
This paper restricts attention to exchangeable models, in particular,
we consider functionals $d^j$ that are independent of the index $j$.
We refer to the above formulation as \textit{joint network inference} (JNI).

\subsection{Network prior} \label{priors}

The network prior [equation~(\ref{Gibbs2})] requires a penalty functional
$d \dvtx \mathcal{G}\times\mathcal{G} \rightarrow\mathbb{R}$ and a prior
network $G^0 \in\mathcal{G}$, with the former capturing how close a
candidate network $G \in\mathcal{G}$ is to the latter.
We discuss choice of $G^0$ below.
Given $G^0$, a simple choice of penalty\vspace{1pt} function $d$ is the structural
Hamming distance (SHD) given by $d(G,G^0) = \| G - G^0 \|$, where $\|
M\| = \sum_{i,j}|m_{i,j}|$ is the $\ell_1$-norm of an adjacency matrix
and the differential network $G - G^0$ is defined to have edges that
occur in exactly one of the networks $G$, $G^0$ [see also \citet{Ibrahim,Imoto}].
The hyperparameter $\eta$ controls the strength of the prior network
$G^0$ [equation~(\ref{Gibbs2})].
Motivated by an application in cancer biology where prior structural
information $G^0$ is available, we follow \citet{Penfold} by restricting
attention to SHD priors, however, our statistical formulation is
general (see below) and compatible with other penalty functionals.
Alternatively, one could employ a beta-binomial prior as described in,
for example, \citet{Dondelinger12}, that allows for the hyperparameters
of the binomial to be integrated out [see also \citet{Oyen}].
Note that in the latter case it is not possible to integrate specific
prior structural information, making beta-binomial priors unsuitable
for the application that this paper considers.

Given a latent network $G$, individual networks $G^j$ are regularized
in a similar way, as $d^j(G^j,G) = \|G^j-G\|$.
In their work on combining multiple data sources, \citet{Werhli} allow
the $\lambda^j$ to vary over individuals $j \in\mathcal{J}$.
Likewise, \citet{Penfold} learn the $\lambda^j$ on a per-individual basis.
However, in both studies, hyperparameter elicitation is nontrivial (see
Section~\ref{elicit}).
In the present paper, we consider only the special case where $\lambda
^1 = \lambda^2 = \cdots= \lambda^J := \lambda$.

A graph $G \in\mathcal{G}$ can be characterized by (i) its adjacency
matrix or
(ii) its parent sets as $G = (\pi_1,\ldots, \pi_P)$, where $\pi_p
\subseteq\mathcal{P} = \{ 1 \cdots P \}$ are the parents of vertex $p$
in $G$.
We write $\mathcal{G}_p$ for the set of possible parent sets for $p$,
such that formally $\mathcal{G} = \mathcal{G}_1 \times\cdots\times
\mathcal{G}_P$.
Although we focus on SHD priors, the inference procedures presented in
this paper apply to the more general class of ``modular'' priors, that
may be factored over $p \in\mathcal{P}$
and written in the form
%
\begin{equation}
d\bigl(G,G^0\bigr) = \sum_{p \in\mathcal{P}}
d_p\bigl(\pi_p,\pi_p^0\bigr),
d^j\bigl(G^j,G;Z^j\bigr) = \sum
_{p \in\mathcal{P}} d_p^j\bigl(
\pi_p^j,\pi_p;Z^j\bigr)
\end{equation}
for some functionals $d_p,d_p^j\dvtx\mathcal{G}_p\times\mathcal{G}_p
\rightarrow\mathbb{R}$.
Here $\pi_p^0$ and $\pi_p^j$ are parent sets for variable $p$,
corresponding to $G^0$ and $G^j$, respectively.

In general, inference for the JNI model [equations~(\ref{Gibbs2}), (\ref{Gibbs1})] will be computationally intensive, as demonstrated in \citet{Werhli,Penfold}.
In Section~\ref{seccomputers} below we show that efficient, \textit{exact} inference is nevertheless possible within the DBN class of
graphical models.

\section{Joint network inference: DBNs} \label{seccomputers}

The JNI model and network priors, as described above, are general. To
apply the JNI framework in a particular context requires an appropriate
likelihood at the individual level, that is, specification of the joint
distribution $p(\mathbf{Y}^j|G^j,\bolds{\theta}^j,Z^j)$ of observables $\mathbf
{Y}^j$ given network $G^j$, ancillary information $Z^j$ and parameters
$\bolds{\theta}^j$, together with a prior distribution $p(\bolds{\theta
}^j|G^j,Z^j)$ over model parameters.
We focus on time-course data, using DBNs and exploiting families of
conjugate prior distributions.
We show that factorization properties of the DBN likelihood permit
computationally tractable joint inference and provide an explicit
algorithm based on belief propagation.

\subsection{DBN formulation}
A DBN is a graphical model based on a DAG on the vertex set $\mathcal
{P}\times\mathcal{T}$, where $\mathcal{T}$ is a set of time indices
[Figure~\ref{DBN}(a); see \citet{Murphy}]. This DAG with $PT$ vertices
is known as the ``unrolled'' DAG.
Here, following \citet{Hill} and others, we use DBNs that permit only
edges forward in time and that
are stationary in the sense that neither the network nor parameters
change with time. For such DBNs, the network can be described by a
directed graph $G$ with exactly $P$ vertices, with edges understood to
go forward in time in the unrolled DAG [see Appendix~\ref{appB} and Figure~\ref{DBN}(b)]. Note that $G$ may have cycles. In what follows, all graphs
(prior, latent and individual)
describe the latter $P$-vertex representation.

Under a modular network prior, structural inference for DBNs can be
carried out efficiently
as described in \citet{Hill}.
In brief, the posterior $G^j|\mathbf{y}$ factorizes into a product of local
posteriors $\pi_p^j|\mathbf{y}$, one factor for each target variable~$p$.
Background and assumptions for DBNs are summarized in Appendix~\ref{appB}; for
general background on DBNs we refer the interested reader to \citet
{Murphy} and for relevant details concerning the class of DBNs used
here to \citet{Hill}.

Write $\mathbf{y}(t)$ for the matrix of observed data at time $t$ for all
individuals $j$ and variables $p$.
In order to simplify notation, we define a data-dependent functional
%
\begin{equation}
\mathfrak{P}(\mathbf{X}) = p\bigl(\mathbf{X}(1)\bigr) \prod
_{t=2}^m p\bigl(\mathbf{X}(t)|\mathbf{y}(t-1)\bigr)
\end{equation}
that implicitly conditions upon observed history. Let $y_p^j(t)$ denote
the observed value of variable $p$ in individual $j$ at time $t$. The
above notation allows us to conveniently summarize the product
%
\begin{equation}
p\bigl(y_p^j(1)|\pi_p^j\bigr)
p\bigl(y_p^j(2)|\mathbf{y}(1),\pi_p^j
\bigr) \cdots p\bigl(y_p^j(m)|\mathbf {y}(m-1),
\pi_p^j\bigr)
\end{equation}
as $\mathfrak{P}(\mathbf{y}_p^j|\pi_p^j)$. Thus, we have that, for DBNs,
the full likelihood also satisfies
%
\begin{equation}\label{factorise}
p\bigl(\mathbf{y}|G^1,\ldots,G^J,Z^1,
\ldots,Z^J\bigr)  =  \prod_{j \in\mathcal{J}} \prod
_{p \in\mathcal{P}} \mathfrak{P}\bigl(\mathbf{y}_p^j|
\pi_p^j,Z^j\bigr),
\end{equation}
where $\mathbf{y}$ denotes the complete data (for all individuals,
variables and times).
In other words, the parent sets $\pi_p^j$ for $p \in\mathcal{P}$, $j
\in\mathcal{J}$ are mutually orthogonal in the Fisher sense, so that
inference for each may be performed separately.

\subsection{Efficient, exact joint estimation} \label{secBMA}
We carry out exact inference in this setting using belief propagation
[\citet{Pearl3}].
Belief propagation is an iterative procedure in which messages are
passed between variables in such a way as to compute exact marginal
distributions; in this respect belief propagation belongs to a more
general class of iterative algorithms known as ``sum-product''
algorithms [\citet{Kschischang}].
Our algorithm is summarized as follows (for simplicity we suppress
dependence upon ancillary information $Z^j$):

\begin{figure}[t]

\includegraphics{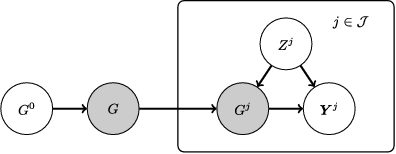}

\caption{Marginalization of JNI over continuous (unknown) parameters
$\bolds{\theta}^j$. (Shaded nodes are unobserved. $G^0={}$prior network,
$G={}$latent network, $G^j={}$network specific to individual $j$, $\mathbf{Y}^j={}$observables for individual $j$, $Z^j={}$ancillary information
available on individual $j$. Hyperparameters $\eta$, $\lambda^j$, $\phi
^j$ are suppressed for clarity. Panel notation is used to indicate the
presence of multiple individuals $j \in\mathcal{J}$.)}
\label{marginalJNIgraph}
\end{figure}

\begin{longlist}[(3)]
\item[(1)] We begin by marginalizing over parameters $\bolds{\theta}^j$ and
caching the local scores $\mathfrak{P}(\mathbf{y}_p^j|\pi_p^j)$ for all
parent sets $\pi_p^j \in\mathcal{G}_p$, all variables $p \in\mathcal
{P}$ and all individuals $j \in\mathcal{J}$; these could be obtained
using any DBN likelihood.
In this paper we exploited conjugate priors to obtain exact expressions
for marginal likelihoods [equation~(\ref{ML}), see Appendix~\ref{appC} for details].
\item[(2)] Following marginalization, the JNI graphical model collapses to
the discrete Bayesian network shown in Figure~\ref{marginalJNIgraph},
whose nodes are themselves graphs.
\item[(3)] Posterior marginal\vspace{1pt} distributions $p(\pi_p|\mathbf{y}_p,\pi_p^0)$ and
$p(\pi_p^j|\mathbf{y}_p,\pi_p^0)$ are then computed using belief
propagation on this discrete Bayesian network.
Pseudocode for this step is provided in Algorithm \ref{alg} in Appendix~\ref{appD}.
\end{longlist}

Let $\mathbf{A}_{\mathrm{JNI}}$ denote the $P \times P$ matrix of marginal
posterior inclusion probabilities for edges in the latent network $G$,
that is,
$(\mathbf{A}_{\mathrm{JNI}})_{ip} := p(i \in\pi_p|\mathbf{y},G^0)$.
These quantities are analogous to posterior inclusion probabilities in
Bayesian variable selection and are computed, using Bayesian model
averaging, as
%
\begin{equation}
(\mathbf{A}_{\mathrm{JNI}})_{ip} = p\bigl(i \in\pi_p|
\mathbf{y},G^0\bigr) = \sum_{\pi_p
\in\mathcal{G}_p}
\mathbf{1}_{i \in\pi_p} p\bigl(\pi_p|\mathbf{y},
\pi_p^0\bigr),
\end{equation}
where $\mathbf{1}_A$ is the indicator of the event $A$ and similarly for
individual networks $(\mathbf{A}_{\mathrm{JNI}}^j)_{ip} := p(i \in\pi_p^j|\mathbf
{y},G^0)$.

Following \citet{Hill}, we reduced the space of parent sets $\mathcal
{G}_p$ using an in-degree sparsity restriction of the form $|\pi_p^j|
\leq c$ for all $\pi_p^j \in\mathcal{G}_p$, $p \in\mathcal{P}$, $j
\in\mathcal{J}$. Thus, the cardinality of the space of parent sets
$|\mathcal{G}_p| = \mathcal{O}(P^c)$ is polynomial in~$P$, where it was
previously super-exponential.
As in variable selection, the bound $c$ should be chosen large enough
that $\mathcal{G}_p$ includes the true data-generating model with high
probability.

Caching of selected probabilities is used to avoid redundant recalculation.
Pseudocode is provided in Algorithm~\ref{alg} in Appendix~\ref{appD}, consisting
of three phases of computation.
Storage costs are dominated by phases I and II, each requiring the
caching of $\mathcal{O}(JP^{1+c})$ terms.
Phase II dominates computational effort, with total (serial)
algorithmic complexity $\mathcal{O}(J^2P^{1+2c})$.
However, within-phase computation is ``embarrassingly parallel'' in the
sense that all calculations are independent (indicated by square
parentheses notation in the pseudocode).
In practice, we have found that problems of size $P \leq20$, $J \leq
20$, $c \leq3$ can be solved within minutes using serial computation
on a standard laptop computer.
We provide serial and parallel MATLAB R2014a implementations in
Supplement B [\citet{suppB}].

\subsection{Network prior elicitation} \label{elicit}
Elicitation of hyperparameters for network priors is an important and
nontrivial issue.
Here we specify the hyperparameters $\lambda,\eta$ in a subjective manner.
We do so due to reported difficulties in estimation of hyperparameters
for related models [\citet{Werhli,Dondelinger12,Penfold}].
We present three criteria below that, for the special case of SHD, are
simple to implement and can be used for expert elicitation.
These heuristics seek to relate the hyperparameters to more directly
interpretable measures of the similarity and difference that they
induce between prior, latent and individual networks:
(i) First, we note the following formula for the probability of
maintaining the status (present/absent) of a candidate parent $i \in
\mathcal{P}$ between the latent network $G$ and an individual network $G^j$:
%
\begin{equation}\label{probinterpret}
h_\lambda:= p\bigl(i \notin\pi_p^j\Delta
\pi_p\bigr) = \frac{1}{1+e^{-\lambda}}.
\end{equation}
This probability provides an interpretable way to consider the
influence of $\lambda$.
For example, a prior confidence of $h_{\lambda} \approx0.73$ that a
given edge status in $G$ is preserved in a particular individual $G^j$
translates into an odds ratio $h_{\lambda}/(1-h_{\lambda}) \approx2.7$
and a hyperparameter $\lambda\approx1$ (see SFigure~1 in the supplementary material\vspace{1pt} [\citet{suppA}]).
An analogous equation relates $\eta$ and $h_\eta:=p(i \notin\pi
_p\Delta\pi_p^0)$, allowing prior strength to be set\vspace{1pt} in terms of the
probability that an edge status in the prior network $G^0$ is
maintained in the latent network $G$.
(ii) A second, related approach is to consider the expected total SHD
between an individual network $G^j$ and the latent network $G$:
%
\begin{equation}\label{totalSHD}
\mathbb{E}\bigl(\bigl\|G^j-G\bigr\|\bigr) = P^2(1-h_\lambda).
\end{equation}
This can be interpreted as the average number of edge changes needed to
obtain~$G^j$ from $G$.
An analogous equation holds for $\eta$ and $h_\eta$.
(iii) Third, in certain applications, the latent network $G$ may not
have a direct scientific interpretation, in which case the criteria
presented above may be unintuitive.
Then, hyperparameters can be elicited by consideration of (a)
similarity between individual networks $G^j,G^k$ and (b) concordance of
individual networks $G^j$ with the prior network $G^0$ (see Supplement A [\citet{suppA}] for further discussion).

\subsection{An information sharing bound}
Below we consider the extent to which information can be shared between
individuals within JNI, providing an upper bound that is attained as
the number of individuals $J$ grows large.
To formalize the contribution to inference from information sharing,
we consider the case in which no data is available on a specific
individual (without loss of generality, individual $j=1$) and
analytically quantify the extent to which JNI can estimate the true
network $\overline{G^1}$ by ``borrowing strength'' from the data $\mathbf
{Y}^2, \ldots, \mathbf{Y}^J$ that represent observations on the remaining
individuals.
(Over-lines will be used to signify the ``true'' data-generating networks.)
As a baseline, write $\mathbf{A}_{0}^j = p(i \in\pi_p^j|\mathbf{Y}^j)$ for the
(naive) estimator that prohibits the sharing of information between individuals.
For simplicity we restrict attention to the case where no network prior
is used ($\eta= 0$), the data-generating hyperparameter $\lambda$ is
known and in-degree restrictions are not in place ($c = P$).
Then, with neither data nor prior information available on individual
$1$, it trivially follows that
%
\begin{equation}\label{baseline}
\mathbb{E}_{\mathbf{Y},\overline{G},\overline{G^1},\ldots,\overline{G^J}|\eta
,\lambda} \biggl[ \frac{\| \mathbf{A}_{0}^1 - \overline{G^1} \|}{P^2} \biggr] = \frac{1}{2},
\end{equation}
where the expectation is taken over all possible data-generating
networks and corresponding data.

From standard, independent network inference we know that consistent
estimation requires unbiasedness of the likelihood function $p(\mathbf
{Y}^j|G^j)$, in the sense that $\mathbb{E}_{\mathbf{Y}^j|\overline{G^j}}
p(\mathbf{Y}^j|G^j)$ is maximized by $G^j = \overline{G^j}$.
We therefore begin by constructing the analogous regularity condition
for joint estimation:
Write $\mathbf{R}$ for the matrix that encodes the prior metric on $\mathcal
{G}$ as $(\mathbf{R})_{G,G'} = \exp(-\lambda\|G-G'\|) / C(\lambda)$, where
$C(\lambda) = \sum_{G \in\mathcal{G}} \exp(-\lambda\|G\|)$. Write $\mathbf
{S}$ for the matrix of expected Bayesian scores $(\mathbf{S})_{G^j,\overline
{G^j}} = \mathbb{E}_{\mathbf{Y}^j|\overline{G^j}} p(\mathbf{Y}^j|G^j)$.

\begin{assumption*}[(Joint regularity)]
For each column of the matrix $\mathbf{M} = (\mathbf{R}\mathbf{S}\mathbf
{R})_{G,\overline{G}}$, the nondiagonal entries are strictly smaller
than the diagonal entry, that is, $M_{G,\overline{G}} < M_{\overline
{G},\overline{G}}$ for all $G \neq\overline{G}$.
\end{assumption*}

To gain intuition for the joint regularity assumption, consider the
special case where $\lambda\rightarrow\infty$; here $\mathbf{R} = \mathbf{I}$
and we only require that the expected local Bayesian score $(\mathbf
{S})_{G^j,\overline{G^j}}$ is maximized by $G^j = \overline{G}$, that
is, we recover the unbiasedness condition from standard network inference.

\begin{theorem*}\label{THM1}
Under the joint regularity assumption, there exists $0 < \varepsilon< 1$
such that
%
\begin{equation}\label{bound}
\mathbb{E}_{\mathbf{Y},\overline{G},\overline{G^1},\ldots,\overline{G^J}|\eta
,\lambda} \biggl[\frac{\| \mathbf{A}_{\mathrm{JNI}}^1 - \overline{G^1}
\|}{P^2} \biggr] = f(J) +
\frac{1}{1+e^{\lambda}},
\end{equation}
where $f(J) \leq2 P^2 \varepsilon^{J-1} \rightarrow0$ as $J \rightarrow
\infty$.
\end{theorem*}

\begin{pf}
See Appendix~\ref{appA}.
\end{pf}

Comparing equation~(\ref{bound}) to  (\ref{baseline}), we see that
information sharing offers gains in estimation, agreeing with
intuition, with larger gains when the true networks are almost
homogeneous ($\lambda$ large).
Moreover, the statistical power of JNI to estimate $\overline{G^1}$
converges to its maximum exponentially quickly as $J\rightarrow\infty$.

\section{Results} \label{secresults}

The proposed methodology was compared against several existing network
inference algorithms. We restricted attention to methods that are
compatible with time-course data and, following the majority of the
literature, carry out \mbox{estimation} for each individual separately.
The computational demands of \citet{Niculescu,Werhli,Penfold} precluded
application in this setting.
Specifically, in the simulated data examples we report below, over 3000
rounds of inference were performed in total, on problems larger than
DREAM4 ($P = 10$, $J = 5$). Using the approach of \citet{Penfold}, these
experiments would have required more than 10 years serial computational
time; in contrast, our approach required less than 24 hours serial
computation on a standard laptop. Thus, we consider the following methods:

\begin{longlist}[(iii)]
\item[(i)] \textit{DBN}. A dynamic Bayesian network, as described in \citet{Hill}, including
nonlinear interaction terms.
For this choice of model it is possible to construct a fully conjugate
set of priors, delivering a closed-form expression for the local
Bayesian score $\mathfrak{P}(\mathbf{y}_p^j|\pi_p^j,Z^j)$.
The model is summarized in Appendix~\ref{appB}.

\item[(ii)] \textit{IDBN}.  \citet{Spencer} recently proposed an extension of
\citet{Hill} that allows analysis of data sets that contain
interventions; this is outlined in Appendix~\ref{appB}.
Interventional DBNs (IDBNs) inherit the computational advantages of
DBNs, in the sense that there is a closed-form expression for the local
Bayesian score, but extend DBNs in a causal direction.
We considered two alternative implementations of IDBNs: (i) \textit{IDBN}.
The approach of \citet{Spencer} was applied to each individual
separately. (ii) \textit{Mono IDBN}. Data on all individuals were pooled
together and fed into a single IDBN analysis, an approach that \citet{Werhli} described as ``monolithic.''

\item[(iii)] \textit{Rel Nets}. A popular approach within the bioinformatics
community is to
score edges based on Pearson correlation of participating nodes
[``relevance networks''; see, e.g., \citet{Butte}]. Here, we used a
time-course analogue in which the correlation
is calculated between successive time points.

\item[(iv)] \textit{LASSO}. An $\ell_1$-penalized likelihood was used to obtain
estimates for coefficients in a linear autoregressive model.
Coefficients were estimated for each \mbox{variable} independently, taking
each variable in turn as the response.
The penalty parameters $\lambda_p$ were each selected using
leave-one-out cross-validation.
Nonzero coefficients indicated the presence of edges.
Further details appear in Supplement~A [\citet{suppA}].
\end{longlist}
Note that DBN and IDBN are able to integrate a prior network $G^0$,
whereas Rel Nets and LASSO are not.
JNI facilitates joint estimation given a suitable graphical model likelihood.
We applied JNI to the DBN and IDBN models described above.
This resulted in several proposed estimators:

\begin{longlist}[(viii)]
\item[(v)] \textit{J-DBN}.  JNI applied to DBN.
\item[(vi)] \textit{J-IDBN}. JNI applied to IDBN.
\item[(vii)] \textit{Fixed IDBN}. Here we formed the likelihood assuming a
single graph for all individuals and the latent network (i.e., $G^1 =
\cdots= G^J = G$) but with parameters allowed to differ. This can be
considered a joint analogue of Mono IDBN that allows
individual-specific parameter values.
\item[(viii)] \textit{AJ-IDBN}. A computationally efficient approximation to
J-IDBN, in which the latent network topology is first estimated using
Fixed IDBN. This is in turn used as an informative network prior within
$J$ independent rounds of IDBN. In this way information sharing is
allowed to occur, but at the expense of a coherent joint posterior.
\end{longlist}

In the empirical study below we compare JNI variants (v)--(viii) against
existing methods (i)--(iv).

\subsection{Performance metrics} \label{metrics}

The proposed methodology addresses three questions, some or all of
which may be of scientific interest depending on the application: (i)
estimation of the latent network $G$, (ii) estimation of individual
networks $G^1,\ldots,G^J$, and (iii) estimation of differences between
individual networks [``differential networks''; \citet{Ideker}].
We quantify performance for each task using the area under the receiver
operating characteristic (ROC) curve (AUR). This metric, equivalent to
the probability that a randomly chosen true edge is preferred by the
inference scheme to a randomly chosen false edge, summarizes, across a
range of thresholds, the ability to select edges in the data-generating network.
AUR may be computed relative to the true latent network $G$ or relative
to the true individual networks $G^j$, quantifying performance on tasks
(i) and (ii), respectively. Both sets of results are presented below, in
the latter case averaging AUR over all individual networks.
For (iii), in order to assess ability to estimate differential
networks, we computed AUR scores based on the statistics
$F_{ip}^j = |p(i \in\pi_p^j|\mathbf{y},G^0,Z^j) - p(i \in\pi_p|\mathbf
{y},G^0,Z^1,\ldots,Z^J)|$ that should be close to one if $i \in\pi
_p^j\Delta\pi_p$, otherwise $F_{ip}^j$ should be close to zero.

It is easy to show that inference for the latent network, under only
the prior (i.e., $\hat{G} = G^0$), attains mean AUR equal to $h_\eta$.
Similarly, prior inference for the individual networks (i.e., $\hat
{G}^j = G^0$) attains mean AUR equal to $1-h_\eta-h_\lambda+2h_\eta
h_\lambda$. This provides a baseline for the proposed methodology at
tasks (i) and (ii) and allows performance to be decomposed into AUR due
to prior knowledge and AUR contributed through inference.

Using a systematic variation of data-generating parameters, we defined
15 distinct data-generating regimes described below.
For all 15 regimes we considered 50 independent data sets; standard
errors accompany average AUR scores.
Results presented below use a computationally favorable in-degree
restriction $c = 3$.
In order to check robustness to $c$, a subset of experiments were
repeated using $c = 4$, with close agreement observed (SFigure 4 in the supplementary material [\citet{suppA}]).

\subsection{Simulation study} \label{silico}

\subsubsection{Data generation}
Data were generated according to DBN models (Appendix~\ref{appB}) as described
in detail in Supplement A [\citet{suppA}].
This data-generating scheme was extended to mimic interventional
experiments that are a feature of our application to breast cancer. In
this case, for each time course, a~randomly chosen variable is marked
as the target of an interventional treatment. Data were then generated
according to the augmented likelihood described in Appendix~\ref{appB} (fixed
effects were taken to be zero).

\subsubsection{Model misspecification and nonlinear data-generating models}
We assume exchangeable networks; it is therefore interesting to explore
the performance of the proposed estimators when the assumption of
exchangeability is \mbox{violated}.
Specifically, we consider a ``worst case'' scenario where individual
networks $G^1, \ldots, G^J$ are sampled from a mixture model with two
distinct components.
Moreover, we consider the extreme case where networks in distinct
mixture components share only a few edges in common; it is expected
that exchangeable estimators will exhibit poor performance in this scenario.
Further, in order to investigate the impact of model misspecification
at the level of the time-series model itself, we considered time-course
data generated from a computational model of protein signaling, based
on nonlinear ODEs [\citet{Xu}].
In order to extend this model, which is for a single cell type, to
simulate a heterogeneous population, we selected three protein species
per individual (at random) and deleted their outgoing edges to obtain
the data-generating networks $G^j$ (see Supplement A [\citet{suppA}]).

\subsubsection{Estimator performance} \label{latent}

We consider the three estimation tasks:
\paragraph*{Latent network} We investigated ability to recover the
latent network $G$.
The existing approaches (i)--(iv) estimate only individual-specific networks.
For estimation of the latent, shared network using these methods, we
simply took an \mbox{unweighted} average of the $J$ estimated adjacency matrices.
The proposed joint estimators (v)--(viii) were assigned hyperparameter
values $\eta= 1, \lambda= 2$ [$\lambda= 3$ for \citet{Xu}] based on
the heuristic of equation~(\ref{probinterpret});
sensitivity to misspecified hyperparameter values is investigated later
in Section~\ref{misspecify}.
Results based on simulated data with interventions are displayed in
STable 3 (see supplementary material [\citet{suppA}]).
We found little difference in the ability of J-IDBN, Fixed IDBN and
AJ-IDBN to recover the latent network structure across a wide range of
regimes, though J-IDBN achieved best performance in 9 out of 15 regimes.
Interestingly, we found that the IDBN estimator, which performs an
unweighted average of $J$ independent inferences, performed
significantly worse than each of J-IDBN, Fixed-IDBN and AJ-IDBN in, respectively, 15, 13 and 11 out of 15 regimes.
Similarly, all the above approaches clearly outperformed Mono IDBN and
Rel Nets, which were in turn outperformed by inference based on the
prior alone, demonstrating the importance of accounting for
individual-specific parameter values.
The joint formulation of DBNs (J-DBN) significantly outperformed
standard DBNs, with higher AUR in all 15 regimes.
LASSO performed best in the regime with long time series ($n=10$) but
failed in other regimes to outperform inference based on the prior alone.
We obtained qualitatively similar results for both alternative
data-generating schemes (STables 4--5, see supplementary material [\citet{suppA}]).

\paragraph*{Individual networks}
At this task, J-IDBN outperformed all other approa\-ches in 9 out of 15 regimes.
AJ-IDBN offered a similar level of performance and together these
estimators demonstrated better performance compared to alternatives in
13 out of 15 regimes.
Since AJ-IDBN avoids intensive computation, this may provide a
practical estimator of individual networks in higher dimensional settings.
Again, the joint approaches J-IDBN and J-DBN both outperformed the
standard approaches IDBN and DBN, respectively, demonstrating an
increase in statistical power resulting from the proposed methodology.
Rel Nets and LASSO performed poorly at this task.
Similar results were observed using the alternative data-generating
schemes (STables 4--5, see supplementary material [\citet{suppA}]).

\paragraph*{Differential networks}
Since JNI regularizes between individuals,
we sought to test whether it could eliminate spurious differences
and thereby improve estimation of differential networks.
Differential networks may also be estimated
using existing methods (i)--(iv); to do so, in each case we compared
individual network estimates with the
estimate of the latent network obtained as described in Section~\ref{latent} above.
We found that, while estimation of differential networks appears to be
more challenging than the other tasks, J-IDBN outperformed the other
approaches in 7 out of 15 regimes.
Moreover, the J-IDBN and J-DBN methods outperformed IDBN and DBN,
respectively, in all 15 regimes.
These results suggest that coherence of joint analysis aids in
suppressing spurious features for estimation of differential network topology.
Rel Nets performed poorly at this task and LASSO performed slightly better.
Intriguingly, AJ-IDBN performed well in estimating differential
networks, performing best in 7 out of 15 regimes.
This suggests that the approximate joint estimator may be suited to
estimation of differential networks.
Results on the noninterventional data sets supported this conclusion
(STable 4, see supplementary material [\citet{suppA}]).
On the \citet{Xu} data sets, however, IDBN and Rel Nets were among the
best performing estimators (STable 5, see supplementary material [\citet{suppA}]), despite being misspecified
for the nonlinear data-generating model.

\subsubsection{Robustness}

We assess three aspects of robustness:

\paragraph*{Hyperparameter misspecification} \label{misspecify}
For the above investigation we used equation~(\ref{probinterpret}) to
elicit hyperparameters $\eta,\lambda$. This was possible because the
data-generating parameters were known by design, however, in general
this will not be the case.
We therefore sought to empirically investigate the effect of
hyperparameter misspecification.
SFigure 3 (see supplementary material [\citet{suppA}]) displays how performance of the J-IDBN estimator for latent
networks depends on the choice of hyperparameters $\lambda,\eta$.
Performance does not appear to be highly sensitive to the precise
hyperparameter values used and there is a large region in which AUR
remains high.

\paragraph*{Outliers and batch effects}
The biological data sets that motivate this study often contain
outliers. At the same time, experimental design may lead to batch effects.
In order to probe estimator robustness, we generated data as described
above, with the addition of outliers and certain batch effects.
Specifically, Gaussian noise from the contamination model $0.95\mathcal
{N}(0,0.1^2) + 0.05\mathcal{N}(0,10^2)$ was added to all data prior to
inference.
At the same time, one individual's data were replaced entirely by
Gaussian white noise to simulate a (strong) batch effect that could
arise, for example, if preparation of a specific biological sample was
incorrect.
The relative decrease in performance at feature detection is reported
in SFigure 5 (see supplementary material [\citet{suppA}]).
We found that J-IDBN remained the best-performing estimator for all
three estimation problems. However, for the differential network
estimation task,
in particular, the decrease in performance
was pronounced for joint methods.

\paragraph*{Nonexchangeability}
SFigure 6 (see supplementary material  [\citeauthor{suppA} (\citeyear{suppA})]) displays the result of inference on data where the
exchangeability assumption is violated.
It can be seen that the performance of all (exchangeable) estimators
decreases in these circumstances, but the magnitude of the decrease is
small (e.g., for estimation of individual networks, J-IDBN experiences
a 0.01 decrease in AUR).
We note that the proposed estimators can be extended to nonexchangeable
settings where elements of the structure that relates individuals are
known; see \citet{Oates4} for further details.

\subsection{Protein signaling networks in breast cancer} \label{secreal}

\begin{figure}

\includegraphics{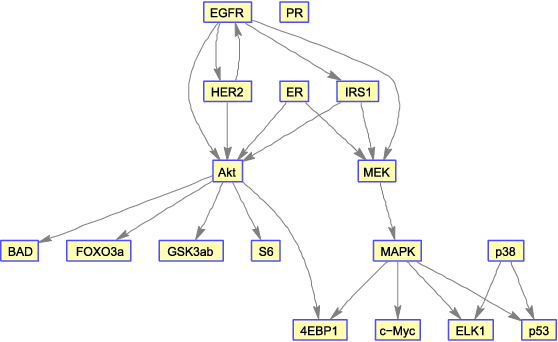}

\caption{Signaling downstream of the epidermal growth factor receptor (EGFR).
The graph shown summarizes known causal links
characterized by extensive biochemistry. (Note that edges in the graph
represent high-level summaries of often complex molecular interactions
that may involve latent chemical species.)}
\label{literature}
\end{figure}

We consider experimental data derived from human breast cancer cell
lines, focusing on protein signaling networks within which many (wild
type) causal relationships are well understood from extensive
biochemistry (Figure~\ref{literature}).
The investigation presented below serves three purposes:
First, it allows investigation of the applicability of the proposed
joint approaches to experimental data.
Second, it allows investigation of the use of ancillary information, in
the form of mutational status and histological information.
Finally, the results and approach are relevant to the topical question
of exploring signaling heterogeneity across cancer cell lines.

Data were obtained using reverse-phase protein arrays [\citet{Hennessy}]
from $J=6$ breast cancer cell lines (AU 565, HCC 1569, MCF 7, MDA MB
231, SKBR3 and SUM 190PT; experimental protocol is described in brief
in Supplement A [\citet{suppA}]).
Data comprised observations for the $P=17$ proteins shown in Figure~\ref{literature} (see also STable 1
in the supplementary material [\citet{suppA}]; we note that these data form part of a
larger study including further cell lines and proteins).
Specifically, $\mathbf{y}$ contains the logarithms of the measured concentrations.
Data were acquired under treatment with an EGFR/HER2 inhibitor
Lapatinib (``EGFRi''), an Akt inhibitor (``Akti''), EGFRi and Akti in
combination, and without inhibition (``DMSO'') at 0.5, 1, 2, 4, 8 and 24
hours following Serum stimulation, giving a total of $n_j=24$
observations of each variable in each individual cell line.

\subsubsection{Informative priors on causal structure} \label{genomics}

For the cancer cell lines analyzed here, ancillary information is
available in the form of genetic aberrations (mutation statuses) and
histological profiling. These were obtained from published sources
[\citet{Neve}] and online databases [\citet{Forbes}] and reproduced in
STable 2 (see supplementary material [\citet{suppA}]).
These sources give causally relevant information on structure specific
to the individual cell lines $j \in\mathcal{J}$. We used this
information to help specify priors on the graphs $G^j$, considering in
particular two cases:
(i) Loss-of-function mutations in kinase domains; in line with the
nature of the mutation, here we set the prior probability on edges
emanating from the mutant protein to zero. Where the mutation is known
to also affect the ability of a protein to be phosphorylated, then
incoming edges were also assigned zero prior probability.
(ii) Cell lines with ectopic expression of the receptor HER2 are known
to depend heavily upon EGFR signaling. In this case the network prior
did not penalize edges emanating from the EGFR receptor nodes.
A~full discussion of ancillary data appears in Supplement A [\citet{suppA}].

In addition to the cell-line-specific mutational information above,
decades of experimental work (including interventional, biochemical and
biophysical studies) have provided a wealth of information about (wild
type) causal relationships between nodes. We used this noncell
line-specific information to specify a prior graph $G^0$ that was
common to all cell lines $j \in\mathcal{J}$ (shown in Figure~\ref{literature}).
Cancer signaling is expected to differ with respect to wild type
signaling, but {a priori} we expect the differences to be small in
number. In light of this observation, we used subjective elicitation
(Section~\ref{elicit}) to set hyperparameters $\lambda= 4,\eta= 5$,
corresponding to $\mathbb{E}(\|G^j-G\|) \approx5$, $\mathbb{E}(\|
G-G^0\|) \approx2$.

\subsubsection{Validation}
\begin{figure}[t]

\includegraphics{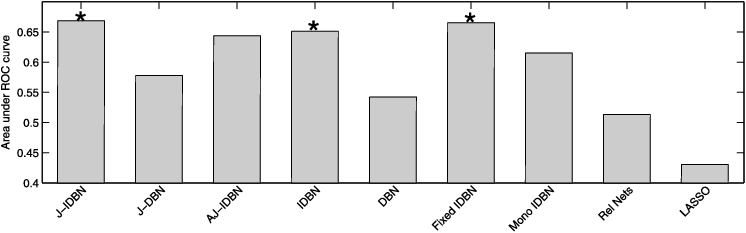}

\caption{Results from breast cancer cell line data, comparison with
network based on literature.
The methods shown were used to estimate a latent network;
AUR is with respect to the literature-based network shown in Figure~\protect\ref{literature}; the latter was not used to provide prior information in
these experiments. (Asterisks denote AUR scores which were significant
at the 1\% level under a permutation test with AUR as the statistic and
$10\mbox{,}000$ samples used to obtain an empirical null distribution.)}
\label{WTAUR}
\end{figure}

In order to test performance, we first considered the latent network
$G$, comparing estimates to the (causal) literature network shown in
Figure~\ref{literature}.
For a fair assessment we used an empty prior network $G^0$.
Inferred networks are displayed in SFigure 7 (see supplementary material [\citet{suppA}]).
Results demonstrated good recovery of the literature network, with
J-IDBN attaining the highest AUR (0.67, $p < 0.01$, permutation test;
Figure~\ref{WTAUR}).
As in the simulation study, J-IDBN outperformed IDBN, with AJ-IDBN and
Fixed IDBN representing good alternative estimators and the remaining
estimators performing poorly.
This suggests the conclusions drawn in Section~\ref{silico} apply also
to the analysis of biological time series data.
In particular, modeling of interventions appears to be crucial in this
setting, in line with the conclusions of \citet{Spencer}.

\begin{figure}[b]

\includegraphics{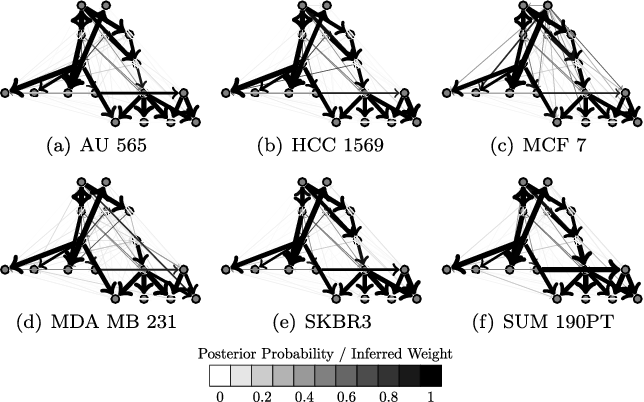}

\caption{Breast cancer data; cell line-specific networks inferred by
J-IDBN. (Edge width and color are proportional to posterior marginal
inclusion probabilities. The layout of vertices is congruent to Figure~\protect\ref{literature}, which can be used as a key.)}
\label{celllinenetworks}
\end{figure}

\subsubsection{Inference for cell line networks}

\begin{figure}[t]

\includegraphics{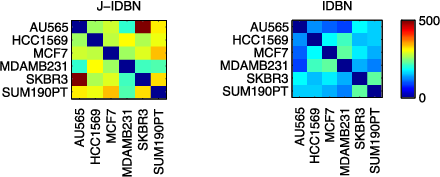}

\caption{Breast cancer data; pairwise similarity between cell
line-specific networks inferred by J-IDBN (left) and IDBN (right). J-IDBN
identifies AU 565 and SKBR 3 as having the most similar networks; these
cell lines were originally derived from the same patient. In contrast,
IDBN does not do so. [Colors denote Bonferroni $-\log(p)$ values based
on the Pearson correlation coefficient of posterior inclusion
probabilities for pairs of cell lines, so that red indicates a high
degree of similarity. For presentation the diagonal is set to zero.]}
\label{pairwise}
\end{figure}

\begin{figure}[b]

\includegraphics{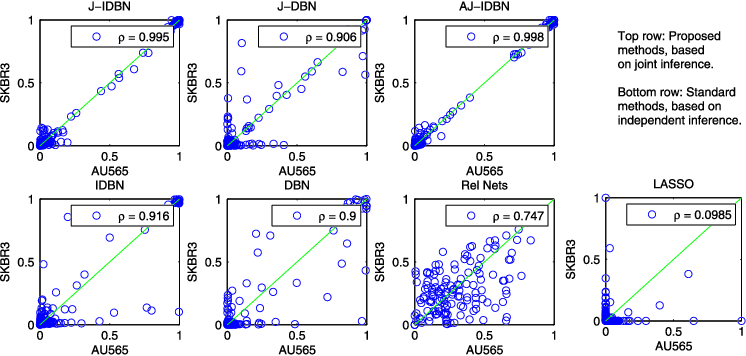}

\caption{Comparison of posterior edge probabilities obtained from
analysis of data from two breast cancer cell lines (AU 565 and SKBR 3)
that were originally derived from the same patient. The joint
estimators J-IDBN and J-DBN improve the Spearman correlation
coefficient (``rho'') between posterior edge probabilities compared to
independent inference using IDBN and DBN.}
\label{samepatient}
\end{figure}

We investigated inference for cell line-specific networks $G^j$ (Figure~\ref{celllinenetworks}), taking the prior network $G^0$ from the
literature (Figure~\ref{literature}).
In order to assess results, we exploited the fact that cell lines AU565
and SKBR3 derive from the same patient.
We would therefore expect these two cell lines to be most similar at
the network level. J-IDBN networks for AU565 and SKBR3 were indeed the
most similar, maximizing the Pearson correlation coefficient between
corresponding posterior marginal inclusion probabilities over all ${6
\choose2} = 15$ pairs of cell lines.
In contrast, standard IDBNs did not do so (Figure~\ref{pairwise}).
Figure~\ref{samepatient} compares posterior inclusion probabilities
(or analogous edge weights for the non-Bayesian methods) for AU565
against SKBR3. We find posterior edge probabilities from these two
lines are closer under JNI estimators compared with standard,
independent estimators.
However, a thorough assessment of the accuracy of the individual cell
line-specific networks requires additional experimental work and is
beyond the scope of this paper.

\section{Discussion} \label{secdiscussion}

We focused on three related structure learning problems arising in the
context of a set of nonidentical but exchangeable units or individuals:

\begin{longlist}[(3)]
\item[(1)] Estimation of a shared network from the heterogeneous data.
\item[(2)] Estimation of networks for specific individuals.
\item[(3)] Learning features specific to individuals (``differential networks'').
\end{longlist}
Each problem may be of independent scientific interest; the joint
approaches investigated here address all three problems simultaneously
within a coherent statistical framework.
We considered simulated data, with and without model misspecification,
as well as proteomic data obtained from cancer cell lines. For all
three problems we demonstrated that a joint analysis performs at least
as well as independent or simpler aggregate analyses.

We considered modular priors (that factorize over nodes) that
facilitated efficient computation. However, it may be useful to
consider richer priors for joint estimation. One possibility that is
pertinent to applications in cancer biology would be hierarchical
regularization that allows entire pathways to be either active or inactive.
However, we note that this would require revisiting hyperparameter
elicitation since the heuristics we described are specific to SHD priors.
We restricted the joint model to have equal inverse temperatures
$\lambda^1 = \cdots= \lambda^J := \lambda$. Relaxing this assumption
may improve robustness to batch effects that target single individuals,
since then weak informativeness ($\lambda^j \approx0$) may be learned
from data. It would also be interesting to distinguish between
$G\setminus G^j$ (``loss of function'') and $G^j\setminus G$ (``gain of
function'') features.
In this work we did not explore information sharing through parameter
values $\bolds{\theta}^j$, yet this may yield more powerful estimators of
network structure in settings where individuals' parameters $\bolds{\theta
}^j,\bolds{\theta}^k$ are not independent.

The case of exchangeable networks that we considered here represents
the simplest of a more general class of models for related networks.
In a sequel to the present paper [\citet{Oates4}],
we discuss the case where multiple individuals are related according to
a known tree structure.
In this more general setting, efficient algorithms based on belief
propagation continue to apply, since the tree constraint ensures that
the corresponding factor graph is acyclic and so the sum-product lemma
continues to hold [\citet{Kschischang}].
Still more general (and challenging) is the case where both the
networks and the hierarchical structure that relate them to one another
are unknown.
\citeauthor{suppA} (\citeyear{suppA,suppB})  present a first step in this direction, in the context of
MAP estimation for nonexchangeable DAGs.

\begin{appendix}

\section{\texorpdfstring{Proof of \protect\hyperref[THM1]{Theorem}}{Proof of Theorem}}\label{appA}

The following \hyperref[lem1]{Lemma} shows that, under the joint regularity assumption,
JNI is a consistent estimator of the true latent network $\overline{G}$
in the limit $J \rightarrow\infty$:

\begin{lemma*}\label{lem1}
Let $\eta= 0$.
Then under the joint regularity assumption there exists $0 < \varepsilon<
1$ such that $\mathbb{E}_{\mathbf{Y},\overline{G},\overline{G^1},\ldots
,\overline{G^J}|\eta,\lambda} p(\overline{G}|\mathbf{Y}) > 1 - |\mathcal
{G}|\varepsilon^J$.
\end{lemma*}

\begin{pf}
Since we are using a flat prior ($\eta= 0$) on $G$, we have,
suppressing dependence upon $\lambda$,
%
\begin{equation}
p(\overline{G}|\mathbf{Y}) = \frac{p(\mathbf{Y}|\overline{G})}{\sum_{G \in
\mathcal{G}} p(\mathbf{Y}|G)},
\end{equation}
so from Jensen's inequality
%
\begin{eqnarray}
\mathbb{E}_{\mathbf{Y},\overline{G^1},\ldots,\overline{G^J}|\overline
{G},\lambda} p(\overline{G}|\mathbf{Y}) & \geq& \frac{\mathbb{E}_{\mathbf
{Y},\overline{G^1},\ldots,\overline{G^J}|\overline{G},\lambda} p(\mathbf
{Y}|\overline{G})}{\sum_{G \in\mathcal{G}} \mathbb{E}_{\mathbf{Y},\overline
{G^1},\ldots,\overline{G^J}|\overline{G},\lambda} p(\mathbf{Y}|G)}
\\
& = & \biggl[ 1 + \mathop{\sum_{G \in\mathcal{G}}}_{G \neq\overline{G}}
\frac{\mathbb{E}_{\mathbf{Y},\overline{G^1},\ldots,\overline
{G^J}|\overline{G},\lambda}p(\mathbf{Y}|G)}{\mathbb{E}_{\mathbf{Y},\overline
{G^1},\ldots,\overline{G^J}|\overline{G},\lambda}p(\mathbf{Y}|\overline{G})} \biggr]^{-1}
\\
& > & 1 - \mathop{\sum_{G \in\mathcal{G}}}_{G \neq\overline{G}}
\frac{\mathbb{E}_{\mathbf{Y},\overline{G^1},\ldots,\overline{G^J}|\overline
{G},\lambda}p(\mathbf{Y}|G)}{\mathbb{E}_{\mathbf{Y},\overline{G^1},\ldots
,\overline{G^J}|\overline{G},\lambda}p(\mathbf{Y}|\overline{G})}
\\
& = & 1 - \mathop{\sum_{G \in\mathcal{G}}}_{G \neq\overline{G}} \prod
_{j \in\mathcal{J}} \frac{\mathbb{E}_{\mathbf{Y}^j,\overline
{G^j}|\overline{G},\lambda}p(\mathbf{Y}^j|G)}{\mathbb{E}_{\mathbf{Y}^j,\overline
{G^j}|\overline{G},\lambda}p(\mathbf{Y}^j|\overline{G})}.
\end{eqnarray}

The joint regularity assumption is equivalent to the requirement
that
$\mathbb{E}_{\mathbf{Y}^j,\overline{G^j}|\overline{G},\lambda} p(\mathbf
{Y}^j|G)$ has a unique maximum at $G = \overline{G}$, since
%
\begin{eqnarray}
\hspace{9pt}\mathbb{E}_{\mathbf{Y}^j,\overline{G^j}|\overline{G},\lambda} p\bigl(\mathbf{Y}^j|G\bigr) & = &
\mathbb{E}_{\overline{G^j}|\overline{G},\lambda} \mathbb{E}_{\mathbf
{Y}^j|\overline{G^j}} \sum
_{G^j \in\mathcal{G}} p\bigl(\mathbf{Y}^j|G^j\bigr) p
\bigl(G^j|G\bigr)
\\
& = & \sum_{\overline{G^j} \in\mathcal{G}} p\bigl(G^j|G\bigr) \sum
_{G^j \in
\mathcal{G}} \bigl[\mathbb{E}_{\mathbf{Y}^j|\overline{G^j}} p\bigl(
\mathbf{Y}^j|G^j\bigr)\bigr] p\bigl(\overline{G^j}|
\overline{G}\bigr)
\\
& = & \sum_{\overline{G^j} \in\mathcal{G}} \sum_{G^j \in\mathcal{G}}
\bigl(\mathbf{R}^T\bigr)_{G,G^j} (\mathbf{S})_{G^j,\overline{G^j}}
(\mathbf{R})_{\overline
{G^j},\overline{G}}
\\
& = & \bigl(\mathbf{R}^T\mathbf{S}\mathbf{R}\bigr)_{G,\overline{G}} =
(\mathbf{R}\mathbf{S}\mathbf{R})_{G,\overline{G}},
\end{eqnarray}
where we have used that $\mathbf{R}$ is symmetric. It follows that
%
\begin{equation}
\varepsilon:= \mathop{\max_{G \in\mathcal{G}}}_{G \neq\overline{G}}
\frac{\mathbb{E}_{\mathbf{Y}^j,\overline{G^j}|\overline{G},\lambda}p(\mathbf
{Y}^j|G)}{\mathbb{E}_{\mathbf{Y}^j,\overline{G^j}|\overline{G},\lambda}p(\mathbf
{Y}^j|\overline{G})} < 1.
\end{equation}
We therefore conclude that
%
\begin{equation} \label{last}
\mathbb{E}_{\mathbf{Y},\overline{G^1},\ldots,\overline{G^J}|\overline
{G},\lambda} p(\overline{G}|\mathbf{Y})  >  1 - |\mathcal{G}|
\varepsilon^J.
\end{equation}
Since equation~(\ref{last}) is independent of $\overline{G}$, the result follows.
\end{pf}

\begin{pf*}{Proof of \hyperref[THM1]{Theorem}}
Since no observables are available on the first individual ($\mathbf{Y}^1 =
\varnothing$), we have
%
\begin{equation}
\mathbf{A}_{\mathrm{JNI}}^1 = \sum_{G \in\mathcal{G}}
p(G|\mathbf{Y}) \sum_{G^1
\in\mathcal{G}} p\bigl(G^1|G
\bigr) G^1.
\end{equation}
We also require the ``oracle'' estimator (O-JNI); this is simply JNI
but with $\overline{G}$ fixed and known, that is,
%
\begin{equation}
\mathbf{A}_{\mathrm{O\aaa JNI}}^1 = \sum_{G^1 \in\mathcal{G}}
p\bigl(G^1|\overline {G}\bigr) G^1.
\end{equation}
Note that $\mathbb{E}_{\overline{\mathbf{G}}|\eta,\lambda} \|\mathbf{A}_{\mathrm{O\aaa JNI}}^1 - \overline{G^1}\| = \mathbb{E}_{\overline{G^1},\overline
{G}|\lambda} \|\overline{G} - \overline{G^1}\| = P^2 (1 - h_{\lambda})$.
We begin by showing that JNI approximates O-JNI:
%
\begin{eqnarray}
\mathbf{A}_{\mathrm{O\aaa JNI}}^1 - \mathbf{A}_{\mathrm{JNI}}^1
& = & \bigl(1 - p(\overline {G}|\mathbf{Y})\bigr) \sum
_{G^1 \in\mathcal{G}} p\bigl(G^1|\overline{G}\bigr)
G^1\nonumber
\\[-8pt]
\\[-8pt]
&& {}- \mathop{\sum_{G \in\mathcal{G}}}_{G \neq\overline{G}} p(G|\mathbf
{Y}) \sum_{G^1 \in\mathcal{G}} p\bigl(G^1|G\bigr)
G^1
\nonumber
\end{eqnarray}
and, by the triangle inequality,
%
\begin{eqnarray}
\bigl\|\mathbf{A}_{\mathrm{O\aaa JNI}}^1 - \mathbf{A}_{\mathrm{JNI}}^1
\bigr\|& \leq& \biggl\llVert \bigl(1 - p(\overline{G}|\mathbf{Y})\bigr) \sum
_{G^1 \in\mathcal{G}} p\bigl(G^1|\overline {G}\bigr)
G^1 \biggr\rrVert
\nonumber
\\
&& {}+ \biggl\llVert \mathop{\sum_{G \in\mathcal{G}}}_{G \neq\overline
{G}}
p(G|\mathbf{Y}) \sum_{G^1} p\bigl(G^1|G
\bigr) G^1 \biggr\rrVert
\\
& \leq& \bigl(1 - p(\overline{G}|\mathbf{Y})\bigr) \sup_{G^1 \in\mathcal{G}}
\bigl\|G^1\bigr\| + \bigl(1 - p(\overline{G}|\mathbf{Y})\bigr) \sup
_{G^1 \in\mathcal{G}}\bigl\|G^1\bigr\|
\\
& \leq& 2\bigl(1 - p(\overline{G}|\mathbf{Y})\bigr) P^2.
\end{eqnarray}
Again, by the triangle inequality,
%
\begin{equation}
\bigl\|\mathbf{A}_{\mathrm{JNI}}^1 - \overline{G^1}\bigr\| \leq\bigl\|
\mathbf{A}_{\mathrm{JNI}}^1 - \mathbf{A}_{\mathrm{O\aaa JNI}}^1
\bigr\| + \bigl\|\mathbf{A}_{\mathrm{O\aaa JNI}}^1 - \overline {G^1}\bigr\|.
\end{equation}
Taking expectations and applying the \hyperref[lem1]{Lemma} produces
%
\begin{equation}
\mathbb{E}_{\mathbf{Y},\overline{\mathbf{G}}|\eta,\lambda} \bigl\|\mathbf{A}_{\mathrm{JNI}}^1 -
\overline{G^1}\bigr\| \leq2 P^2 |\mathcal{G}|
\varepsilon^{J-1} + P^2 (1 - h_{\lambda}),
\end{equation}
as required.
\end{pf*}

\section{Dynamic Bayesian networks}\label{appB}

\begin{figure}

\includegraphics{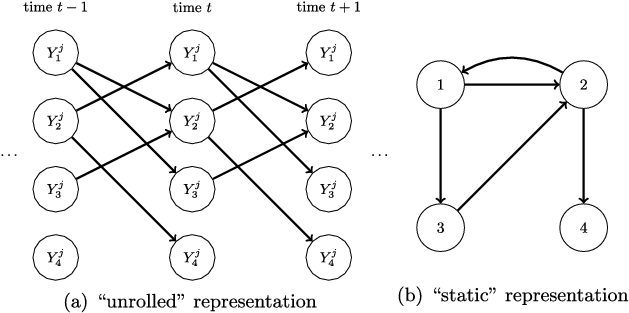}

\caption{Dynamic Bayesian networks (DBNs). \textup{(a)} An ``unrolled'' dynamic
Bayesian network (DBN) showing each variable at successive time points.
\textup{(b)} The corresponding ``static'' representation of DBN \textup{(a)} with exactly
one vertex for each variable.}
\label{DBN}
\end{figure}

For the DBNs used here, an edge $(p,q)$ from $p \in\mathcal{P}$ to $q
\in\mathcal{P}$ in $G^j \in\mathcal{G}$ implies that $Y_q^j(t)$, the
observed value of variable $q$ in individual $j$ at time $t$, depends
directly upon $Y_p^j(t-1)$, the observed value of $p$ in individual $j$
at time $t-1$ [Figure~\ref{DBN}(a); note that $t$ indexes the
sample index rather than actual sampling time].
Let $\mathbf{Y}^j$ denote a vector containing all observations for
individual $j$.
Then $\mathbf{Y}^j(t)$ is conditionally independent of $\{\mathbf{Y}^j(t-\tau
):\tau\geq2\}$ given $\mathbf{Y}^j(t-1)$, $\bolds{\theta}^j$, $G^j$ and $Z^j$
(first-order Markov assumption).
These conditional independence relations are conveniently summarized as
a (static) network $G^j$ with exactly $P$ vertices [Figure~\ref{DBN}(b)]; note that this latter network need not be acyclic.

\citet{Hill} describe a DBN rooted in the Bayesian linear model.
Specifically, the response $Y_p^j(t)$ is predicted by covariates $\mathbf
{Y}^j(t-1)$, that is,
%
\begin{equation}
\mathbf{Y}_p^j = \mathbf{X}_0\bolds{\alpha}
+ \mathbf{X}_{\pi_p^j}^j\bolds{\beta} + \bolds {\varepsilon},
\end{equation}
where $\bolds{\varepsilon} \sim N(\mathbf{0}_{n\times1},\sigma^2\mathbf{I}_{n \times n})$.
In many cases multiple time series will be available. In this case the
vector $\mathbf{Y}_p^j$ contains the concatenated time series.
The matrix $\mathbf{X}_0 = [\mathbf{1}_{\{t=1\}} \enskip   \mathbf{1}_{\{t>1\}}]_{n \times
2}$ contains a term for the initial time point in each experiment.
The elements of $\mathbf{X}_{\pi_p^j}^j$ corresponding to initial
observations $Y_p^j(1)$ are simply set to zero.
Parameters $\bolds{\theta}_p^j = \{\bolds{\alpha},\bolds{\beta},\sigma\}$ are
specific to model $\pi_p^j$, variable $p$ and individual $j$.
In the simplest case, given data $\mathbf{Y} = \mathbf{y}$, the model-specific
component $\mathbf{X}_{\pi_p^j}^j$ of the design matrix consists of the raw
predictors $\mathbf{y}_{\pi_p^j}^j(t-1)$, where $\mathbf{y}_Z^j$ denotes the
elements of the vector $\mathbf{y}^j(t-1)$ belonging to the set $A$, though
more complex basis functions may be used, including interaction terms.
For experiments performed in this paper, interaction terms were taken
to be all possible products of parent variables, following \citet{Hill}.

\citet{Spencer} modeled interventional data by modification to the DAG
using ideas from causal inference [\citet{Pearl}].
We mention briefly some of the key ideas and refer the interested
reader to the references for full details. A ``perfect intervention''
corresponds to 100\% removal of the target's activity with 100\% specificity.
In the context of protein phosphorylation, kinases may be intervened
upon using chemical agents.
\citet{Spencer} make the simplifying assumptions that these
interventions are perfect [the ``perfect out fixed effects'' (POFE) approach].
We refer the reader to \citet{Spencer} for an extended discussion of
POFE. This changes the DAG structure to model the intervention and also
estimates an additional fixed effect parameter to model the change
under intervention in the log-transformed data.
When generating data for the simulation study in Section~\ref{silico}
we take fixed effects to equal zero.

\section{Exact marginal likelihood for DBN and~IDBN} \label{appC}
\citet{Hill} employed an exact Bayesian approach to capture the
suitability of the candidate parent set $\pi_p^j$.
In brief, a Jeffreys prior $p(\bolds{\alpha},\sigma|\pi_p^j,\phi^j,\break Z^j)
\propto1/\sigma$ for $\sigma>0$ was placed over the common parameters.
Prior to inference, the noninterventional components of the design
matrix are orthogonalized using the transformation $(\mathbf{X}_{\pi
_p^j}^j)_{ik} \mapsto\sum_l (\mathbf{I}_n-\mathbf{P}_0)_{il} (\mathbf{X}_{\pi
_p^j}^j)_{lk}$, where $\mathbf{P}_0 = \mathbf{X}_0(\mathbf{X}_0^T\mathbf{X}_0)^{-1}\mathbf
{X}_0^T$ [\citet{Bayarri}].
A $g$-prior was placed on regression coefficients [\citet{Zellner}],
given by
%
\begin{equation}
\bolds{\beta}|\bolds{\alpha},\sigma,\pi_p^j,
\phi^j,Z^j \sim N\bigl(\mathbf{0}_{b
\times1},
\phi^j\sigma^2\bigl(\mathbf{X}_{\pi_p^j}^T
\mathbf{X}_{\pi_p^j}\bigr)^{-1}\bigr),
\end{equation}
where $b = \dim(\bolds{\beta})$.
Using these priors alongside either DBNs or IDBNs as outlined above,
the marginal likelihood can be obtained in closed-form:
%
\begin{eqnarray}
\label{ML} && \mathfrak{P}\bigl(\mathbf{y}_p^j|
\pi_p^j,\phi^j,Z^j\bigr)\nonumber
\\[-8pt]
\\[-8pt]
&& \qquad \propto\frac{1}{(\phi^j+1)^{b/2}} \biggl(\mathbf{y}_p^{jT} \biggl(
\mathbf{I}_{n\times
n}-\mathbf{P}_0-\frac{\phi^j}{\phi^j+1}
\mathbf{P}_{\pi_p^j} \biggr)\mathbf {y}_p^j
\biggr)^{-({n-a})/{2}},\nonumber
\end{eqnarray}
where $\mathbf{P}_{\pi_p^j} = \mathbf{X}_{\pi_p^j}(\mathbf{X}_{\pi_p^j}^T\mathbf{X}_{\pi
_p^j})^{-1}\mathbf{X}_{\pi_p^j}^T$, $a = \dim(\bolds{\alpha})$ and $b = \dim
(\bolds{\beta})$.
Empirical investigations have previously demonstrated good results for
network inference based on the above marginal likelihood [\citet{Hill,Spencer}].

The hyperparameter $\phi^j$, that is related to the weight of the
parameter prior $p(\bolds{\beta}|\bolds{\alpha},\sigma)$ relative to the data
$\mathbf{y}_p^j$, was selected in this paper using the conditional
empirical Bayes procedure outlined in \citet{George}, corresponding to
%
\begin{equation}
\hat{\phi}^j\bigl(\pi_p^j\bigr) = {\arg
\max}_g \mathfrak{P}\bigl(\mathbf{y}_p^j|
\pi_p^j,g,Z^j\bigr).
\end{equation}
For computational efficiency, we evaluated the argument over a set of
eight candidate values corresponding to prior weights of
0, 10, 20, 30, 40, 50\% and $(100/n)\%$ (the unit information prior).
Alternative strategies for eliciting $g$-priors are discussed in \citet{Bayarri,Liang}.

\section{Belief propagation for JNI}\label{appD}

Exact inference for JNI is based on belief propagation [\citet{Pearl}].
Algorithm~\ref{alg} displays pseudocode for exact joint model averaging.
We also indicate computational complexity in terms of the number $M =
|\mathcal{G}_p|$ of possible parent sets and the number $J$ of individuals.
Computational complexity of calculating marginal likelihoods $\mathfrak
{P}(\mathbf{y}_p^j|\pi_p^j)$ will partly depend upon sample size $n$;
scaling exponents shown here assume $\mathcal{O}(n) = \mathcal{O}(1)$.
Algorithm \ref{alg} contains pseudocode for computation of posterior
marginal inclusion probabilities for edges in both the latent network
$G$ and individual-specific networks $G^j$.
For simplicity, we suppress dependence upon ancillary data $Z^j$ throughout.

\begin{algorithm}[b]
\caption{Belief propagation for JNI}
\label{alg}
\fontsize{9.6pt}{13pt}\selectfont{\begin{algorithmic}[1]
\For{$p \in\mathcal{P}$}
\item[Phase 0:]
\State Compute and cache $\mathfrak{P}(\mathbf{y}_p^j|\pi_p^j)$ [$\forall j
\in\mathcal{J}$] [$\forall\pi_p \in\mathcal{G}_p$]
\item[Phase I:]
\State Compute and cache [$\forall j \in\mathcal{J}$] [$\forall\pi_p
\in\mathcal{G}_p$]
\State$\mathfrak{P}(\mathbf{y}_p^j|\pi_p) = \sum_{\pi_p^j \in\mathcal
{G}_p} \mathfrak{P}(\mathbf{y}_p^j|\pi_p^j) p(\pi_p^j|\pi_p)$ [$\mathcal{O}(M)$]
\item[Phase II:]
\State Compute and cache\vspace*{2pt} [$\forall j \in\mathcal{J}$] [$\forall\pi_p,
\pi_p^j \in\mathcal{G}_p$]
\State$p(\pi_p|\mathbf{y}_p,\pi_p^0) \propto p(\pi_p|\pi_p^0) \prod_{j \in
\mathcal{J}} \mathfrak{P}(\mathbf{y}_p^j|\pi_p)$ [$\mathcal{O}(J)$]
\State$p(\pi_p^j|\mathbf{y}_p,\pi_p^0) \propto\sum_{\pi_p \in\mathcal
{G}_p} p(\pi_p|\pi_p^0) \mathfrak{P}(\mathbf{y}_p^j|\pi_p^j) p(\pi_p^j|\pi
_p) \prod_{k \in\mathcal{J}\setminus\{j\}} \mathfrak{P}(\mathbf{y}_p^k|
\pi_p) [\mathcal{O}(MJ)$]
\item[Phase III:]
\State Compute and cache [$\forall j \in\mathcal{J}$] [$\forall i \in
\mathcal{P}$]
\State$p(i \in\pi_p|\mathbf{y},G^0) = \sum_{\pi_p \in\mathcal{G}_p}
\mathbf{1}_{i \in\pi_p} p(\pi_p|\mathbf{y}_p,\pi_p^0)$ [$\mathcal{O}(M)$]
\State$p(i \in\pi_p^j|\mathbf{y},G^0) = \sum_{\pi_p^j \in\mathcal{G}_p}
\mathbf{1}_{i \in\pi_p^j} p(\pi_p^j|\mathbf{y},\pi_p^0)$ [$\mathcal{O}(M)$]
\EndFor
\end{algorithmic}}
\end{algorithm}
\end{appendix}

\section*{Acknowledgments}
We are grateful to the Editor and anonymous referees for feedback that
has improved the content and presentation of this paper.
We would also like to thank J. D. Aston, F. Dondelinger, C. A. Penfold,
S. E. F. Spencer and S. M. Hill for helpful discussion and comments.

%
\begin{supplement}[id=suppA]
\sname{Supplement A}
\stitle{Additional results and protocols\\}
\slink[doi]{10.1214/14-AOAS761SUPPA} 
\sdatatype{.pdf}
\sfilename{AOAS761\_suppa.pdf}
\sdescription{Includes: Alternative data generating models; robustness
to in-degree restriction, outliers, batch effects and
nonexchangeability; ancillary information for breast cancer; inferred
wild type networks for breast cancer.}
\end{supplement}

\begin{supplement}[id=suppB]
\sname{Supplement B}
\stitle{Computational implementation\\}
\slink[doi]{10.1214/14-AOAS761SUPPB} 
\sdatatype{.zip}
\sfilename{AOAS761\_suppb.zip}
\sdescription{MATLAB R2014a code (serial and parallel) implementing
joint network inference.}
\end{supplement}


%



%

\printaddresses
\end{document}